\documentclass[5p,preprint,sort&compress]{elsarticle}

\usepackage{amsmath}
\usepackage{graphicx}
\usepackage{amssymb}
\usepackage{float}
\usepackage{mathpazo}
\usepackage{lineno}
\usepackage[pdftex]{color}
\usepackage{version}
\usepackage{verbatim}
\usepackage{lineno}
\usepackage{cancel}

\usepackage{color}

\pdfoutput=1
\newcommand{\ignore}[1]{}

\newcommand{\be}{\begin{equation}}
\newcommand{\ee}{\end{equation}}
\def \ba#1\ea{\begin{align}#1\end{align}}
\newcommand{\bit}{\begin{itemize}}
\newcommand{\eit}{\end{itemize}}

\renewcommand{\i}{$\mathrm{i}$}

\def \slashb#1{\setbox0=\hbox{$#1$}#1\hskip-\wd0\dimen0=5pt\advance
        \dimen0 by-\ht0\advance \dimen0 by\dp0\lower0.5\dimen0\hbox
          to\wd0{\hss \sl/\/ \hss}}

\input{epsf}

\begin{document}

\title{
Why matter effects matter for JUNO
\vglue -2.0cm \hglue 15cm
{\small FERMILAB-PUB-490-T}
\vglue 2.0cm 
}
\author[AK,SP]{Amir N.\ Khan}
\ead{amir.khan@mpi-hd.mpg.de, akhan@fnal.gov}
\address[AK]{Max-Planck-Institut f\"{u}r Kernphysik, Postfach 103980,
D-69029 Heidelberg, Germany}
\author[HN]{Hiroshi Nunokawa}
\ead{nunokawa@puc-rio.br}
\address[HN]{Departamento de F\'{\i}sica, Pontif\'{i}cia Universidade Cat\'{o}lica do
Rio de Janeiro, C. P. 38097, 22451900, Rio de Janeiro, Brazil }
\author[SP]{Stephen J. Parke}
\ead{parke@fnal.gov}
\address[SP]{Theoretical Physics Department, Fermi National Accelerator Laboratory, Batavia, IL 60510, USA}

\begin{abstract}
In this paper we focus on the Earth matter effects for 
the solar parameter determination by a medium baseline reactor experiment such as JUNO.
We derive perturbative expansions for the mixing angles $\theta_{12}$
and $\theta_{13}$ as well as the $\Delta m^2_{21}$ and $\Delta m^2_{31}$ in terms of 
the matter potential relevant for  JUNO.
These expansions, up to second order in the matter potential, while
simple, allow one to calculate the electron antineutrino survival
probability to a precision much better than needed for the JUNO experiment.
We use these perturbative expansions to semi-analytically explain and confirm the shift caused by the matter effects on
the solar neutrino mixing parameters $\theta_{12}$ and  $\Delta m^2_{21}$ which were previously obtained by a purely numerical $\chi^2$ analysis.  Since these shifts do not
satisfy the naive expectations and are significant given
the precision that can be achieved by the JUNO experiment, a totally independent  cross check using a completely different method  is of particular importance. We find that  these matter effect shifts do not depend on any of the details of the detector characteristics apart from the baseline and earth mass density between reactor(s) and detector, but do depend on the normalized product of reactor neutrino spectrum times the inverse-beta decay cross-section. 
The results of this manuscript suggests an alternative analysis method for measuring  $\sin^2 \theta_{12}$ and  $\Delta m^2_{21}$ in JUNO which would be a useful cross check of the standard analysis and for the understanding of the Wolfenstein matter effect.
The explanation of these shifts together with a quantitative understanding, using a semi-analytical method, is the principal purpose of this paper.  

\end{abstract}

\date{\today }

\maketitle

\section{Introduction}
\label{sec:Introduction}

Jiangmen Underground Neutrino Observatory (JUNO) \cite{JUNOYB2016} is a medium baseline
reactor antineutrino experiment with $\sim$ 50 km kilometer baseline that is
currently under construction.  The primary stated goal of this experiment is to determine the neutrino mass
ordering, i.e. whether the neutrino mass eigenstate with least $\nu_e$
is the most massive, normal ordering or least massive, inverted ordering. 
One of the secondary goals of this experiment
is to measure the solar neutrino mixing parameters ($\sin ^{2}\theta _{12}$ and 
$\Delta m_{21}^{2}$) and the atmospheric $\Delta m^2$'s  with sub percent precision.
In this paper we address the Wolfenstein matter effects on 
the measurement of the solar parameters by JUNO.

To be more precise, we consider the JUNO-like experiment, or
the reactor experiment which has the same main features of JUNO 
in terms of its size, baseline and energy resolution with some
simplifications and approximations 
regarding the experimental setup and/or analysis. 
For brevity, we refer such an experiment simply as JUNO in this paper.
There is a proposal of similar experiment,
RENO-50~\cite{Kim:2014rfa}, but in practice, currently, JUNO is the only 
reactor experiment with medium baseline which is 
expected to start taking data soon, within the next few years.
Therefore, we focus on the JUNO experiment in this paper.

We first derive a simple perturbative expansion for the mixing
angles and $\Delta m^2$'s in matter. Using these matter mixing angles
and matter $\Delta m^2$'s one can calculate 
the electron antineutrino survival probability with a precision that is beyond what is needed for the JUNO experiment.  The fractional precision is better than $10^{-6}$. The matter effects only significantly impact the measurements of 
the solar parameters $\sin ^{2}\theta _{12}$ and $\Delta m_{21}^{2}\ $.  
Previous attempts to approximate the oscillation probability 
relevant for experiment like JUNO
including matter effects~\cite{FCap2014, FCap2015,Yu2016} 
seem to be either very complicated or 
less accurate than the ones shown in
this paper.
The matter mixing angles and mass-squared differences derived here are simple expansions in the matter potential up to second order.

Due to small matter effect, the effective mixing angle and mass squared
difference in matter are slightly shifted from the ones in vacuum,
if they are obtained by ignoring {\it intentionally} the matter
effect.
Certainly, the matter effect should be included when JUNO analyzes its
reactor neutrino data, but this can be performed in a number of different ways.  Using the oscillation probabilities in matter is one important way, another is to use the vacuum oscillation  probabilities and apply the shifts calculated in this manuscript.   It will be an interesting and useful exercise for  JUNO to perform the analyses in both ways and compare them given the understanding provided by  this manuscript on the matter effect shifts.
While it would not be possible to identity or establish 
the presence of the matter effect by JUNO
alone, one can perform the above as a consistency check.

In \cite{Yu2016}, using a full blown numerical $\chi^2$
analysis, it was shown that the fractional shift for the solar
parameters 
due to the Earth matter effect (assuming the constant matter density 2.6
g/cm$^3$) are 
\begin{eqnarray}
\left. \left( \frac{\delta( \sin^2\theta_{12})
 }{\sin^2\theta_{12}}, \frac{ \delta( \Delta m^2_{21}) }{\Delta m^2_{21}} \right) \right|_{\text{purely numerical} ~\chi^2} & & \nonumber \\[3mm]
&&  \hspace{-3.5cm} \simeq 
( -1.1,\quad  0.19)\%.  %\nonumber %
\label{eq:shift_chisq1}
\end{eqnarray}
On the other hand, by considering only the first order in matter
effect, from eqs.~(11) and (17) of ~\cite{Yu2016}, 
we naively expect that the shift
 at fixed neutrino energy, $E$, are
\begin{eqnarray}
 \left. \left( \frac{\delta( \sin^2\theta_{12})
 }{\sin^2\theta_{12}}, \frac{ \delta( \Delta m^2_{21}) }{\Delta
 m^2_{21}} \right) \right|_{ \rm naive\ expectation} & & \nonumber \\[2mm]
& & \hspace{-6.5cm}
\simeq  \left\{\begin{array}{lll}
(-1.1,~~ &0.30)\% & {\rm ~using} ~E=3 {\rm~MeV}, \\[2mm]
(-0.74, &0.21)\% & {\rm ~using} ~E=2 {\rm~MeV}.
\end{array}  \right.  %\nonumber
\label{eq:shift_naive1}
\end{eqnarray}
Comparing these results it is clear that the naive expectation can
explain 
either the shift for $\sin^2 \theta_{12}$ or $\Delta m^2 _{21}$ but not
both at the same time using a fixed neutrino energy. 
That is, in the  ($\sin^2 \theta_{12}$,  $\Delta m^2 _{21}$) plane, the results of ref. \cite{Yu2016} do not lie on the line given by  the naive expectation at any value of the neutrino energy.
We have checked that including also 
higher order matter corrections does not help this situation.
In this paper we explain in a semi-analytical way
the above apparent discrepancy 
and confirm that our predictions are consistent with
the results of purely numerical $\chi^2$ analysis of  \cite{Yu2016}.
For our explanation only the following details of the experimental setup are important, the reactor neutrino spectrum times the inverse-beta decay cross-section and the baseline and matter earth density between the reactor(s) and detector. All other details of the experimental setup are unimportant for the determination of the 
size of the matter effect shifts for  $\sin^2 \theta_{12}$ and $\Delta m^2 _{21}$.

In section 2, we give the expansion for the mixing angles, $\sin^2 \theta$'s and $\Delta m^2$'s as expansions in the matter potential which are used to calculate the electron anti-neutrino survival probability  with sufficient precision for the JUNO experiment. 
In section 3, we discuss the event rates both in vacuum and in matter 
taking into account the smearing for the reconstructed neutrino energy
due to finite energy resolution.
In section 4 we give the naive estimate of the shift in  $\sin^2
\theta_{12}$ and $\Delta m^2_{21}$ as well as details of our semi-analytic estimate
of the shift using the reactor spectral information
as well as the energy dependence of the inverse beta decay cross section.  
Section 5 provides the summary and conclusions.

\section{Electron Neutrino Survival Probability including Matter Effects}
\label{sec:Probability}
The electron neutrino oscillation survival probability in vacuum can be easily derived as
\begin{eqnarray}
P_{ee} &=&1-\cos ^{4}\theta _{13}\sin ^{2}2\theta _{12}\sin^{2}\Delta_{21} 
 \nonumber \\
&& \hskip -1.5cm -\sin ^{2}2\theta _{13}(\cos ^{2}\theta _{12}\sin ^{2}\Delta _{31}+\sin
^{2}\theta _{12}\sin ^{2}\Delta _{32}),  
\label{probvac}
\end{eqnarray}%
where $\Delta _{ij}\equiv \Delta m_{ij}^{2}L/4E$ 
with $\Delta m_{ij} \equiv m_i^2-m_j^2$,
and $m_i (i=1,2,3)$ being neutrino masses,
and we will use the relation that 
$\Delta m^2 _{32}\equiv \Delta m^2_{31}-\Delta m^2_{21}$.

For constant matter density, 
we simply need to replace all the mixing angles and mass-squared differences by the
corresponding matter parameters ($\Delta \widetilde{m}^2_{ij}$) 
and mixing angles ($\widetilde{\theta }_{ij}$), as
\begin{eqnarray}
\widetilde{P}_{ee}
&=&1-\cos ^{4}\widetilde{\theta }_{13}\sin ^{2}2\widetilde{%
\theta }_{12}\sin ^{2}\widetilde{\Delta }_{21}\nonumber \\
&& \hskip -1.5cm -\sin^{2}2\widetilde{\theta }
_{13}[\cos ^{2}\widetilde{\theta }_{12}\sin ^{2}\widetilde{\Delta }%
_{31}+\sin ^{2}\widetilde{\theta }_{12}\sin ^{2} {\widetilde \Delta
}_{32}],  
\label{probmat}
\end{eqnarray}%
where $\widetilde{\Delta }_{ij}\equiv \Delta \widetilde{m}_{ij}^{2}L/4E$.
The mass-squared differences ($\Delta \widetilde{m}_{ij}^{2}$) and mixing
angles ($\widetilde{\theta }_{ij}$) in matter can be calculated using
the exact complicated expressions found in \cite{Zaglauer:1988gz}.

Here we will use the approximate expressions derived by Denton, Minakata
and Parke (DMP)~\cite{DMP1} 
which depend on two quantities, first
the Wolfenstein matter potential times neutrino energy,
\begin{equation}
a\equiv \pm 2\sqrt{2}G_{F}N_{e}E,
\label{eq:matter-potential}
\end{equation}
where the +(-) sign refers to the case where neutrino (antineutrino)
channel is considered,
$G_F$ is the Fermi constant, $N_e$ is the electron number density, 
and second,
\begin{equation}
\Delta m_{ee}^{2} \equiv \cos ^{2}\theta _{12}\Delta m_{31}^{2}+\sin
^{2}\theta _{12}\Delta m_{32}^{2}.
\end{equation}
Since we are interested in reactor neutrinos to be observed by 
JUNO, in this work we will mainly consider antineutrino 
channel (considering the negative sign in
eq.~(\ref{eq:matter-potential})) unless otherwise stated. 

First for $\widetilde{\theta }_{13}$, we have~\cite{DMP1}
\begin{eqnarray}
\cos 2\widetilde{\theta }_{13} & \simeq &
\frac{\Delta m_{ee}^{2}\cos 2\theta _{13}-a%
}{\Delta \widetilde{m}_{ee}^{2}},\text{ }  \label{2th13}
\end{eqnarray}
where 
\begin{eqnarray}
\Delta \widetilde{m}_{ee}^{2}
\equiv\ 
\Delta m_{ee}^{2}\sqrt{(\cos 2\theta
_{13}-a/\Delta m_{ee}^{2})^{2}+\sin ^{2}2\theta _{13}}.  \label{dmee} 
\end{eqnarray}

%%%%%%%%%%%%%%%%%%%%%%%%%%%%%%%%%%%%%%%%%%%%%%%%%%%%%%%%%%%%%%%%%%%%%%%%
\begin{figure*}[h!]
%\vglue -0.5cm
\begin{center}
\includegraphics[width=6.5in]{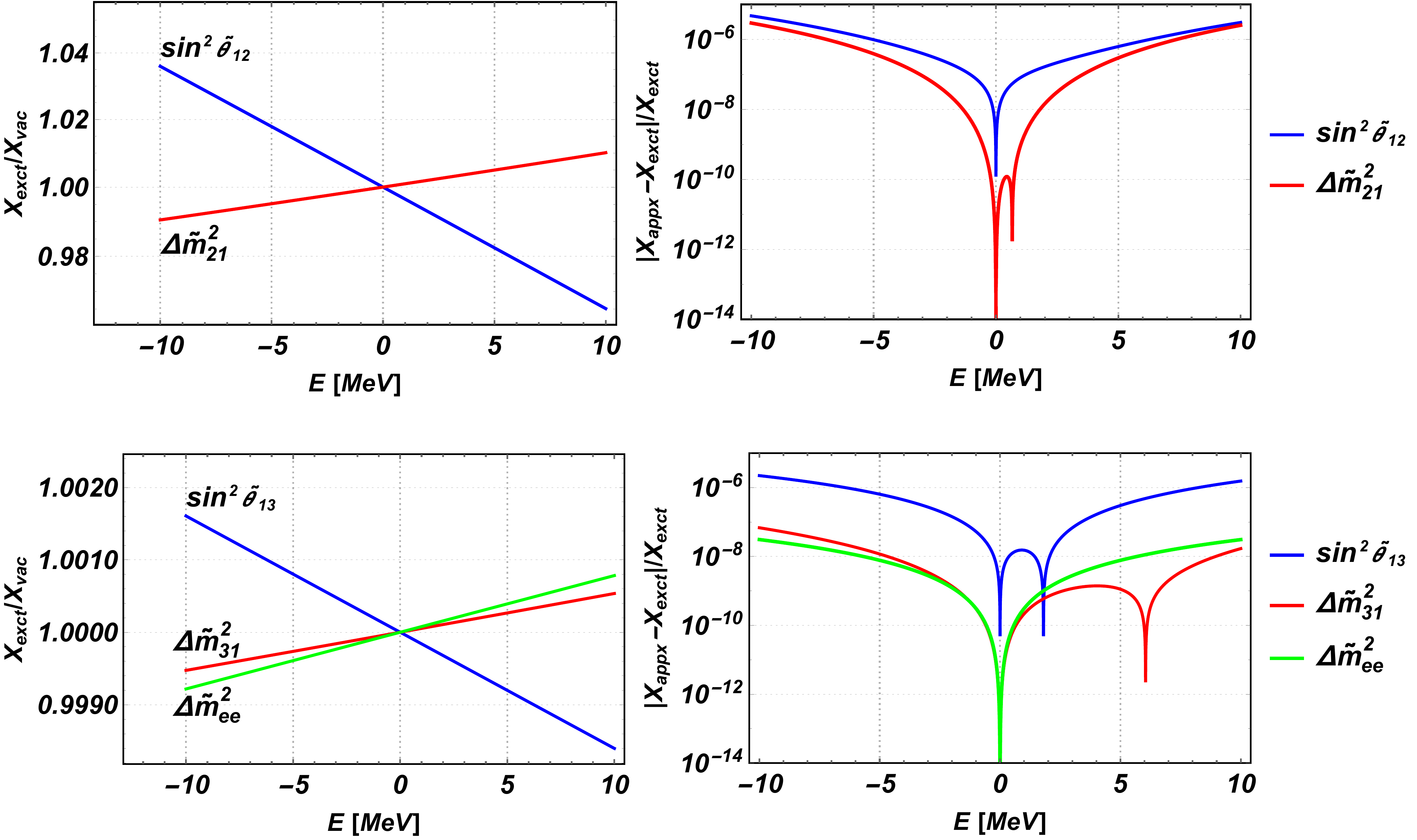}
\end{center}
\vglue -0.5cm
\caption{
In the left panels we show the ratios of all the mixing parameters
which are relevant for medium baseline reactor neutrino experiment, 
in matter and vacuum, 
or ``exact matter/vacuum'', 
computed using the exact formulas in matter~\cite{Zaglauer:1988gz} 
as a function of neutrino energy.
In the right panels we show the fractional error 
between exact and approximated values of mixing parameters in matter, or
``$\vert$ approximated matter - exact matter $\vert$ /exact matter'',
as a function of neutrino energy
where eqs.~(\ref{eq:theta12_matt})-(\ref{eq:dm2_32_matt})
based on DMP results were used for the approximated ones.
We note that the positive (negative) energy corresponds to the
antineutrino (neutrino) channel since as long as the matter effect is
concerned, changing the sign of energy is equivalent to changing the sign
of the potential in eq.~(\ref{eq:matter-potential}). 
The difference between exact and either of the approximations would be invisible on the left panels.}
\end{figure*}
%%%%%%%%%%%%%%%%%%%%%%%%%%%%%%%%%%%%%%%%%%%%%%%%%%%%%%%%%%%%%%%%%%%%%%%%

%%%%%%%%%%%%%%%%%%%%%%%%%%%%%%%%%%%%%%%%%%%%%%%%%%%%%%%%%%%%
\begin{table*}[h!]
\begin{center}
\begin{tabular}{l|l|l|l|l|l|l}
\hline \hline
parameters$\rightarrow $ & $\sin ^{2}\theta _{12}$ & $\sin ^{2}\theta _{13}$
& $\Delta m_{21}^{2}[e$V$^{2}]$ & $\Delta m_{31}^{2}[eV^{2}]$ & Baseline & 
earth matter density \\ \hline
Values$\rightarrow $ & $3.04\times 10^{-1}$ & $2.14\times 10^{-2}$ & $%
7.34\times 10^{-5}$ & $2.455\times 10^{-3}$ & $52.5\ $km & $\rho =$ $2.6$
g/cm$^{3}$ \\ \hline
error (1$\sigma $ \%) & $4.4$ & $3.8$ & $2.2$ & $1.4$ & $-$ & $-$ \\ 
\hline \hline
\end{tabular}%
\end{center}
\vglue -0.4cm
\caption{True (input) parameter values assumed in throughout this work 
unless otherwise stated. We use 0.5  for the number of electrons per nucleon in the earth.
The oscillation parameters values and their errors were taken from
ref. \protect \cite{lisi}. }
\label{tabosc}
\end{table*}
%%%%%%%%%%%%%%%%%%%%%%%%%%%%%%%%%%%%%%%%%%%%%%%%%%%%%%%%%%%%

\clearpage

\clearpage
Then $\widetilde{\theta }_{12}$ and $\Delta \widetilde{m}_{21}^{2}$
\begin{eqnarray}
\cos 2\widetilde{\theta }_{12} &
\simeq  &\frac{\Delta m_{21}^{2}\cos 2\theta
_{12}-a^{^{\prime }}}{\Delta \widetilde{m}_{21}^{2}},  \label{2th12} 
\end{eqnarray}
where
\begin{eqnarray}
\Delta \widetilde{m}_{21}^{2} &
\simeq \ &\Delta m_{21}^{2} \times \nonumber \\
&&\hskip -2.5cm 
\sqrt{(\cos 2\theta
_{12}-a^{^{\prime }}/\Delta m_{21}^{2})^{2}+\cos ^{2}(\widetilde{\theta }%
_{13}-\theta _{13})\sin ^{2}2\theta _{12}}\ , \label{dm21} 
\label{eq:dm2_21_matt1}
\end{eqnarray}
and  the effective matter potential for 1-2 sector, $a^\prime$, is given by
\begin{eqnarray}
a^{^{\prime }} & \equiv & (a+\Delta m_{ee}^{2}-\Delta
 \widetilde{m}_{ee}^{2})/2\,,
\end{eqnarray}
as well as 
\begin{eqnarray}
\cos ^{2}(\widetilde{\theta }%
_{13}-\theta _{13})  & \equiv & \nonumber \\
&&\hskip -2.5cm
(\Delta \widetilde{m}_{ee}^{2}+
\Delta m_{ee}^{2}-a\cos 2\theta _{13})/(2\Delta \widetilde{m}%
_{ee}^{2})\,,  
\end{eqnarray}
and finally $\Delta \widetilde{m}_{31}^{2}$ and $\Delta \widetilde{m}_{32}^{2}$;
\begin{eqnarray}
\Delta \widetilde{m}_{31}^{2} &=& \Delta \widetilde{m}_{ee}^{2}+ \sin^2\widetilde{\theta }_{12} \,\Delta \widetilde{m}_{21}^{2}\,,  \\
\Delta \widetilde{m}_{32}^{2} & =  & \Delta \widetilde{m}_{31}^{2} -  \Delta \widetilde{m}_{21}^{2}.
\end{eqnarray}

Performing a Taylor series expansion in the parameters $a/\Delta m_{ee}^{2}$ 
$\ $and $a/\Delta m_{21}^{2}$ on 
eqs.~(\ref{2th13})-(\ref{eq:dm2_21_matt1}) one can
calculate all the oscillation parameters up to the required accuracy level. In the
following, we write down the mass-squared differences and the mixing angles
in the parameters up to the order accurate enough for JUNO:%:
\begin{eqnarray}
\sin ^{2}\widetilde{\theta }_{12} &
\simeq\ & s_{12}^{2} \nonumber  \\
&& \hskip -2.5cm \times \left[1+2c_{12}^{2}\left(\frac{%
c_{13}^{2}a}{\Delta m_{21}^{2}}\right)+3c_{12}^{2}\cos 2\theta _{12}\left(%
\frac{c_{13}^{2}a}{\Delta m_{21}^{2}}\right)^2 \, \right],   
\label{eq:theta12_matt}  \\[2mm]
\Delta \widetilde{m}_{21}^{2} &
\simeq \ &  \Delta m_{21}^{2}  \nonumber \\
&&\hskip -2.5cm   \times \left[1- \cos 2 \theta_{12} 
\left(\frac{c_{13}^{2}a}{\Delta m_{21}^{2}} \right)%
+2s_{12}^{2}c_{12}^{2} \left(\frac{c_{13}^{2}a}{\Delta
			m_{21}^{2}}\right)^{2}   \right], 
 \label{eq:dm2_21_matt}
\\[2mm]
\sin ^{2}\widetilde{\theta }_{13}  & 
\simeq \  & s_{13}^{2} ~\left[1+2c_{13}^{2}
\left(\frac{a}{\Delta m_{ee}^{2}} \right) \right],  
 \label{eq:theta13_matt} \\[2mm]
\Delta \widetilde{m}_{ee}^{2} &
\simeq &\Delta m_{ee}^{2} ~\left[1- \cos 2 \theta_{13} 
\left(\frac{a}{%
\Delta m_{ee}^{2}}\right) \right],  \label{eq:dm2_ee_matt}
\\[2mm]
\Delta \widetilde{m}_{31}^{2} &
\simeq  & \Delta m_{31}^{2}
 \label{eq:dm2_31_matt} \\[2mm]
&&\hskip -3 cm \times  \left[ 1-  \left(\frac{a}{\Delta m_{31}^{2}}\right)
		  \left((c_{12}^{2}c_{13}^{2}  - s_{13}^{2})   
-  s^2_{12} c^2_{12} c^2_{13}  \left( \frac{c_{13}^{2}a}{\Delta
				m_{21}^{2}}\right) \right) 
     \right],   \nonumber  \\[2mm]
\Delta \widetilde{m}_{32}^{2} & 
\simeq
&\Delta m_{32}^{2}  
 \label{eq:dm2_32_matt} \\[2mm]
& & \hskip -3cm  \times \left[ 1-  \left(\frac{a}{\Delta m_{32}^{2}}\right) 
\left((s_{12}^{2}c_{13}^{2}  - s_{13}^{2})   
+  s^2_{12} c^2_{12} c^2_{13}  \left( \frac{c_{13}^{2}a}{\Delta m_{21}^{2}}\right) \right)      \right]. 
\nonumber 
\end{eqnarray}

Note that the solar parameters, $\sin^2 \widetilde{\theta }_{12}$  and $\Delta \widetilde{m}^2_{21}$, are given to second order in $a/\Delta m^{2}_{21}$ whereas the atmospheric parameters  $\sin^2 \widetilde{\theta }_{13}$  and $\Delta \widetilde{m^2}_{ee}$ are given to only 
first order in $a/\Delta m^{2}_{ee}$.  However, the solar parameters use $\Delta m^2_{21}$, whereas the atmospheric parameters use $\Delta m^2_{ee}$.
Numerically, 
\begin{eqnarray}
\hskip -0.6cm 
\frac{c^2_{13} a}{\Delta m_{21}^{2}} \simeq 
2.55 \times 10^{-2} 
 ~~\frac{(E/10\ \text{MeV}%
)(\rho /2.6 \text{ g cm}^{-3})}{(\Delta m_{21}^{2}/7.34\times 10^{-5}\ \text{eV}%
^{2}) } \,, % \\
\end{eqnarray}
\begin{eqnarray}
\hskip -0.6cm \frac{a}{\Delta m_{ee}^{2}} \simeq 
7.97 \times 10^{-4}~~ \frac{(E/10\ \text{MeV}%
)(\rho /2.6\text{\ g cm}^{-3})}{(\Delta m_{ee}^{2}/2.4\times 10^{-3}\ \text{eV}%
^{2})},
\end{eqnarray}
for $\sin^2\theta_{13}$ given in Table 1,
so that  at  $E \sim 11$ MeV 
\begin{equation}
\left(\frac{c^2_{13} a}{\Delta m_{21}^{2}}%
\right)^{2} \approx \frac{a}{\Delta m_{ee}^{2}},
\end{equation}
which is just beyond the highest energy of reactor neutrinos.

%%%%%%%%%%%%%%%%%%%%%%%%%%%%%%%%%%%%%%%%%%%%%%%%%%%%%%%%%%%%

\begin{figure}[t!]
\vglue 0.5cm
\begin{center}
\includegraphics[width=3.5in]{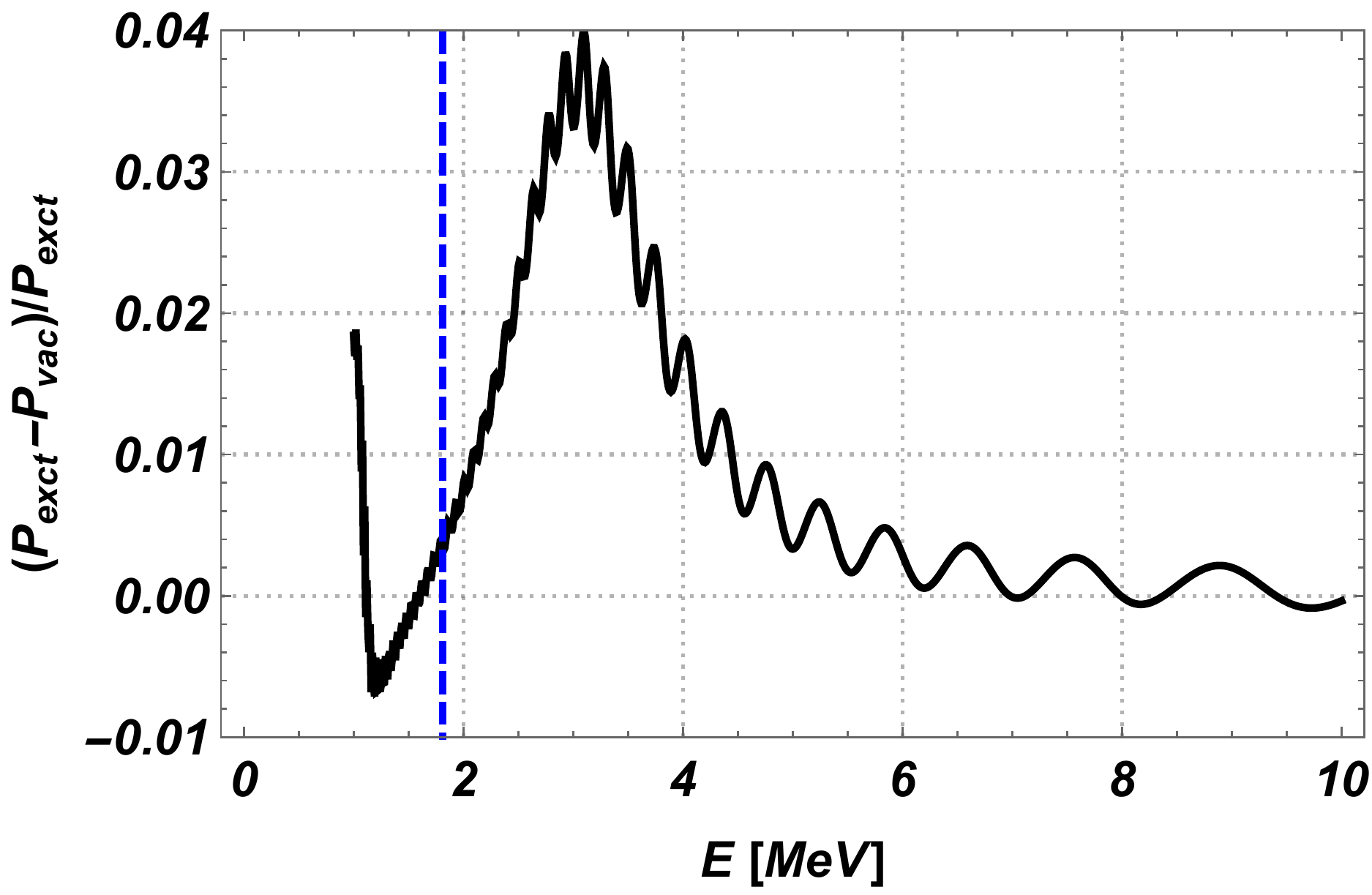}
\end{center}
\vglue -0.5cm
\caption{
The fractional difference of the exact oscillation
probability in matter and the vacuum oscillation probability versus the antineutrino
energy. The maximum matter effects is $\sim$ 4\% around 3 MeV. }
\label{fig:fig2}
%\vglue -0.2cm
\end{figure}
%%%%%%%%%%%%%%%%%%%%%%%%%%%%%%%%%%%%%%%%%%%%%%%%%%%%%%%%%%%%
\begin{figure}[h!]
\begin{center}
%\vglue -0.6cm
\vglue 0.5cm
\includegraphics[width=3.5in]{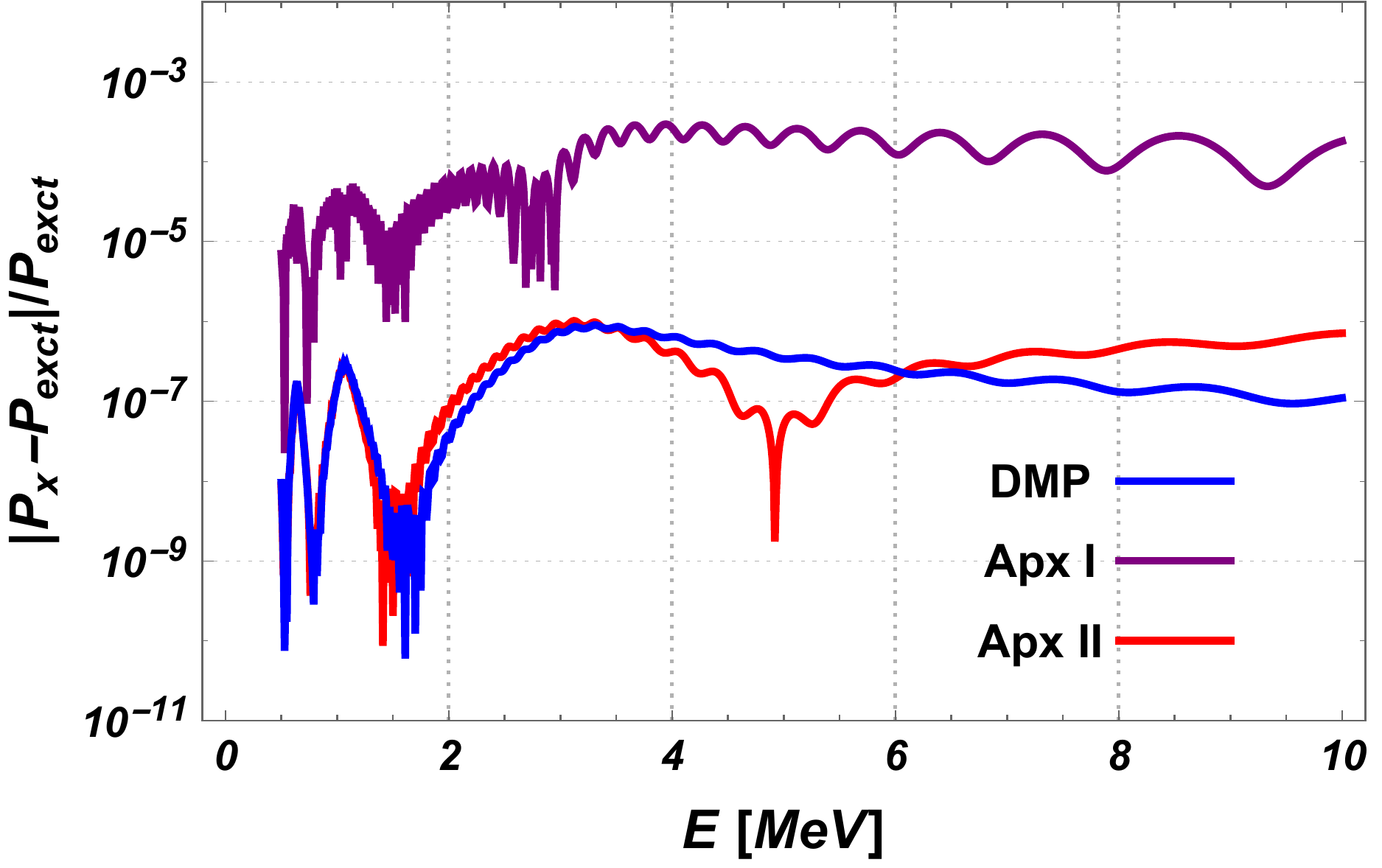}
\end{center}
\vglue -0.6cm
\caption{Absolute fractional difference between the exact and approximated
probabilities in matter versus the antineutrino energy. 
The red (blue) is the difference between our 
approximation obtained by eq.~(\ref{probmat}) 
with eqs.~(\ref{eq:theta12_matt})-(\ref{eq:dm2_32_matt})
(results obtained by DMP formulas~\cite{DMP1}) and the exact result.
We also show by the solid purple curve the same quantity but
computed using only the first order in matter effect.
}
\vglue -0.3cm
\end{figure}
%%%%%%%%%%%%%%%%%%%%%%%%%%%%%%%%%%%%%%%%%%%%%%%%%%%%%%%%%%%%

In Fig. 1 we show the matter effects on each parameter relevant for JUNO
using the mixing parameters shown in Table \ref{tabosc}.
In left panels we show the ratio of the mixing parameters in matter and 
the corresponding ones in vacuum as a function of 
neutrino energy whereas in the right panels
we show the fractional precision given by our approximation 
in eqs.~(\ref{eq:theta12_matt})-(\ref{eq:dm2_32_matt}).
We note that positive (negative) energy corresponds to the case of 
antineutrino (neutrino) channel 
since as long as the matter effect is concerned, changing the sign
of neutrino energy is equivalent to changing that of the matter potential,
see eq.~(\ref{eq:matter-potential}).

Clearly the matter effect on solar parameters is more than 
an order of magnitude larger than for the atmospheric parameters.
The fractional precision figures, right panels of Fig. 1, show that the fractional difference between our approximated
expressions of eqs. (\ref{eq:theta12_matt})-(\ref{eq:dm2_32_matt})
and the exact values are less than a few
$10^{-6}$ for $E$ below 10 MeV.

In Fig. 2 we show the fractional difference
between survival probability in matter 
and that in vacuum versus the antineutrino energy.
The figure clearly shows that the most important region is between 2 to 5 MeV and the
peak occurs around 3 MeV. 
This regime is also the most important for the precision measurement
of the solar oscillation parameters.   

The accuracy of our approximate probability using eq. (\ref{probmat})
with effective mixing parameters in 
eqs.~(\ref{eq:theta12_matt})-(\ref{eq:dm2_32_matt})
for the energy range relevant for the reactor spectrum and  JUNO
baseline is shown in Fig. 3. 
We took the
absolute difference of our approximate probability
with the exact probabilities using the exact oscillation probabilities of
ref.~\cite{Zaglauer:1988gz}.
For comparison, we also show the difference between the
full DMP probability \cite{DMP1} against the exact probability in blue
color.  
Again from Fig. 3, using the second order expansions for the neutrino parameters in matter, our approximate matter oscillation probability has a fractional precision of better than $10^{-6}$   for $E<$ 10 MeV.

In addition, we show by the solid purple curve the same quantity but
computed using only the first order in matter effect.
We note that roughly speaking, if we consider only the statistical error, 
JUNO is expected to measure $\bar{\nu}_e$ survival probabilities 
with an error of $\sim O(0.1)$ \% which is just a factor of 
few times larger than the errors 
of the probability with the first order in matter effect around
4 MeV.

\section{Event Rates}

For the purpose of this paper, we
consider the JUNO detector with the exposure corresponding to 
$(36\times 20\times 5)$\,GW$\cdot$kton$\cdot$years of
reactor power times target mass for five years of running time. 
We compute the differential event rate in terms of the reconstructed neutrino
energy, $E_{\text{rec}}$, as%
\begin{eqnarray}
\frac{dN}{ dE}_{\text{rec}} % (E_{\text{rec}})
&=&
\frac{N_{p}T}{4\pi L^{2}} \int_{m_n-m_p+m_e} dE %
\frac{d\phi }{dE} \nonumber \\  
&& \hskip -2cm \times \widetilde{P}_{ee}(L,E)\sigma _{IBD}(E)
G(E-E_{\text{rec}},\delta E),  \label{eq:spect}
\end{eqnarray}%
where $N_p$ is the total number of target protons in the liquid scintillation
detector of JUNO, $L$ is the baseline, $T$ is the total run time, 
${d\phi }/{dE}$ is the reactor flux we take from ref. 
\cite{Mueller:2011nm}, 
$\widetilde{P}_{ee}(L,E)$ is the
oscillation probability, $\sigma _{IBD}$
is the total cross-section of inverse beta decay at the detector taken from
ref. \cite{Strumia:2003zx}. 
For simplicity and as a good approximation we ignore the small
variation (of the order of $\sim 0.5$\%) of 
the distances from the JUNO detector to the 10 reactor
cores and set $L=52.5$ km.
Despite that we are studying the tiny effect of matter potential,
since we are interested to estimate only the matter induced shift of the solar parameters 
(not the accuracy of the measurement), 
we believe that taking into account the variation of baselines 
have essentially no impact on our results. This is expected to be
true because we treat both the input and fit exactly in the same way 
(by applying the same approximations to both input and fit). 

In eq.~(\ref{eq:spect}) $G(E-E_{\text{rec}},\delta E)$ is the
normalized Gaussian smearing function which takes into account the photon energy smearing of the 
detector. We define this function to be,
\begin{equation}
G(E-E_{\text{rec}},\delta E)=\frac{1}{\sqrt{2\pi }\, \delta E}\exp \left[ -%
\frac{(E-E_{\text{rec}})^{2}}{2(\delta E)^{2}}\right] ,  \label{eq:G}
\end{equation}%
where $\delta E$ is the energy   detector resolution, defined as~\cite{JUNOYB2016} 
\begin{equation}
\frac{\delta E}{E-0.8\ \text{MeV}}
= 3\%\biggr/  \sqrt{\frac{E-0.8\ \text{MeV}}{\text{MeV}}
}\ ,
\label{eq:sigE}
\end{equation}%
following the analysis done in \cite{Abrahao:2015rba}.
Here, we have used that $m_n-m_p-m_e = 0.8$ MeV, 
so that the visible energy deposited in the detector 
is related to the reconstructed neutrino energy 
as $E_\text{vis} \sim E_\text{rec} - 0.8$ MeV
which at threshold is $\sim 1$ MeV.
%Here w
We do not consider neither the proton recoil in the IBD reaction 
nor the so called
``non-stochastic'' terms in eq.~(\ref{eq:sigE}) 
(see eq. (2.11) and related description in \cite{JUNOYB2016})
due to the same reason mentioned before. 
As long as we are interested only in the shift due to matter potential, 
we believe that these details are not important.
For our purpose, we are satisfied by the fact that
by our simplified $\chi^2$ analysis  we could reproduce rather well 
the results shown in Fig. 4 of \cite{Yu2016}.

\begin{figure}[h!]
\begin{center}
\hskip -0.2cm
\includegraphics[width=3.5in]{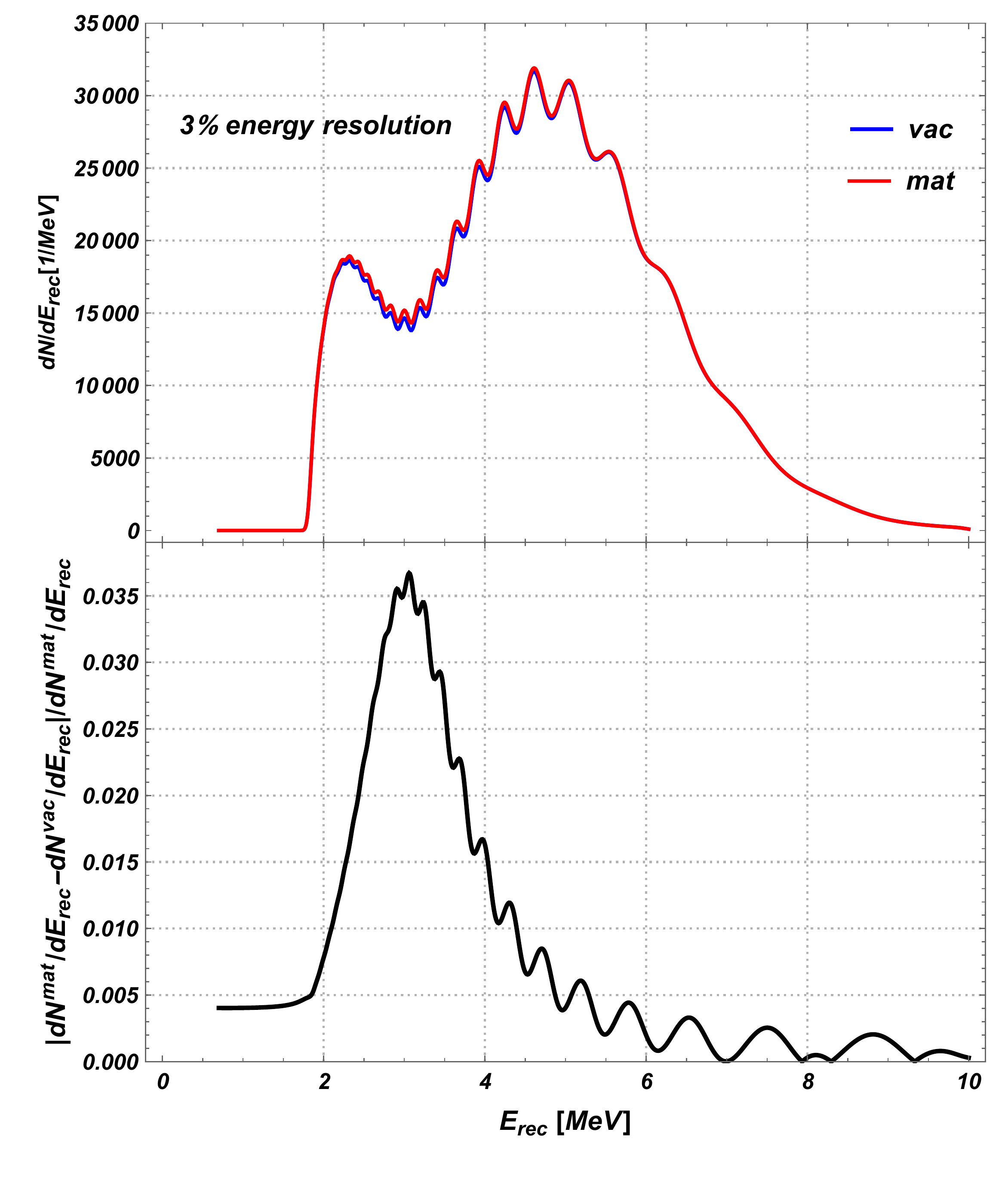}
\end{center}
\vglue -0.9cm
\caption{
Top panel: Spectrum of events at JUNO in terms of the reconstructed neutrino
energy, $E_{\text{rec}}$, 
with the corresponding exposure of 180 GW$\cdot$years
(reactor thermal power times running years) which give 
$\sim 1.1\times 10^5$ events.
The red line uses the oscillation probability in matter, eq.~(\ref{probmat})
while the blue line uses the vacuum oscillation probability.
Bottom panel: 
The fractional difference of the event rates between matter
and vacuum. Note, the maximum appears around 3 MeV, as expected from  Fig.~\ref{fig:fig2}.}
\label{fig:event-distribution}
\vglue -0.2cm
\end{figure}

In the upper panel of Fig.~\ref{fig:event-distribution} we show the 
event number distribution. 
On the lower panel, we show the 
fractional difference of the event rates between matter
and vacuum. 

%%%%%%%%%%%%%%%%%%%%%%%%%%%%%%%%%%%%%%%%%%%%%%%%%%%%%%%%%%%%%
%%%%%%%%%%%%%%%%%%%%%%%%%%%%%%%%%%%%%%%%%%%%%%%%%%%%%%%%%%%%%

\newpage

\section{
Shift of solar parameters, 
$\sin ^{2}\protect \theta _{12}$ and $\Delta m_{21}^{2}$,  due to the
 matter effect at JUNO
}
\label{sec:shift}

In this section we give a precise, quantitative estimate of
the size of the shift due to matter effects for $\theta_{12}$ 
and $\Delta m^2_{21}$ for JUNO using the perturbative expressions given in Sec. \ref{sec:Probability}.  
Namely, we will see how much the results differ between the two cases 
when matter 
effect is taken into account or ignored, by considering
more analytic details than the discussions given in~\cite{Yu2016}
regarding the shift. 
Considering JUNO's expected sensitivity, the matter effect 
should definitely be taken into account and the analysis should be performed in multiple ways as a consistency check.

%%%%%%%%%%%%%%%%%%%%%%%%%%%%%%%%%%%%%%%%%%%%%%%%%%%%%%
We first try to estimate the shift due to the matter 
effect by using analytic probability formulas 
considering only the first order in matter effect
to compute the impact of matter effect for 
a given neutrino energy. 
We will show that by choosing appropriately the
neutrino energy ($\sim$ 3 MeV) we can predict, without performing a fit to the data, 
rather accurately the size of the shift due to the matter effect 
for the mixing angle $\sin ^{2}\protect \theta _{12}$ 
but we tend to overestimate somewhat the shift 
for $\Delta m_{21}^{2}$.
 Alternatively, if we choose $\sim$2 MeV which is at the very low end of
 the observed neutrino spectrum, we can predict the shift in  $\Delta
 m_{21}^{2}$ but underestimate the shift in $\sin ^{2}\protect \theta
 _{12}$. 
As mentioned in the introduction,
including also higher order matter corrections does not help this situation.

Later in this section, we will resolve this apparent discrepancy in a relatively simple manner, by accurately computing
or predicting the shift due to the matter effect 
for  $\sin ^{2}\protect \theta _{12}$  and  $\Delta m_{21}^{2}$, without performing a full $\chi^2$
analysis.  Thus, confirming the full $\chi^2$ analysis in an independent
way
as long as the shift is concerned.

\subsection{Naive estimation of the shift due to matter effect}
First let us try to estimate analytically the expected shift due to
matter effect in the determination of the solar parameters 
$\sin ^{2}\protect \theta _{12}$ and $\Delta m_{21}^{2}$ by JUNO.
For the sake of illustration and simplicity, we consider 
only the first order term in the matter effect in eqs.~(\ref{eq:theta12_matt})
and (\ref{eq:dm2_21_matt}), and adopt the approximation that the rapid oscillation
driven by the atmospheric mass squared difference is simply averaged out, 
since as long as the estimation of the shift due to matter effect 
for solar parameters is concerned, this rapid oscillation 
is  unimportant.

Under this approximation, from eq.~(\ref{probmat}) we can obtain
the survival probability with matter effect which is essentially reduced to
that for
2 flavor system with some small corrections due to non-zero $\theta_{13}$, 
which can be expressed as 
\begin{equation}
\tilde{P}_{ee}(E) 
\simeq 1-{}c^4_{13} \sin ^{2}2 \tilde{\theta}_{12}
\sin ^{2}\left( \frac{\Delta \tilde{m}^{2}_{21}L}{4E}\right) 
-  \frac{1}{2}\sin^2 2\theta_{13},
\label{eq:p_ee-matter-approx}
\end{equation}
where 
$\tilde{\theta}_{12}$  and $\Delta \tilde{m}^{2}_{21}$ 
are given, respectively, through the following expressions, 
\begin{eqnarray}
\sin^2 \tilde{\theta}_{12} &
\simeq  & 
\sin^2 \theta_{12} \left[1-
2\alpha (\rho) E\, \frac{\cos^2
 \theta_{12}}{\cos2 \theta_{12}}\right],
\label{eq:theta12_matt2}
\end{eqnarray}
or equivalently, 
\begin{eqnarray}
\sin ^{2}2\widetilde{\theta}_{12}&
\simeq  &\sin ^{2}2\theta_{12} \lbrack 1-2\alpha
(\rho )E],
\label{eq:theta12_matt3}
\end{eqnarray}
and
\begin{eqnarray}
\Delta \widetilde{m}^{2}_{21} &\simeq &\Delta m^{2}_{21}[1+\alpha (\rho )E],
\label{eq:dm2_matt2}
\end{eqnarray}
with
\begin{eqnarray}
\alpha (\rho ) &\equiv & 
\frac{2\sqrt{2}G_{F}N_{e} c^2_{13}
\cos 2\theta_{12} }{\Delta m^{2}_{21}}%
\approx \nonumber \\
&&\hskip -1cm (1.059\times 10^{-3})\times\, c^2_{13}
\left( \frac{\cos 2 \theta_{12}}{0.392}\right) \nonumber \\
&&\hskip -2cm\left( \frac{\Delta m^2_{21}}{7.34\times 10^{-5}\ \text{eV}^2}\right)^{-1}
\left( \frac{\rho }{2.6 \text{ g cm}^{-3}}\right)\,
\text{MeV}^{-1},
\label{eq:alpha}
\end{eqnarray}
which are obtained by taking into account only the first order
in matter effect in eqs.~(\ref{eq:theta12_matt}) and (\ref{eq:dm2_21_matt}).
We note that shift indicated in 
eqs.~(\ref{eq:theta12_matt3}) and (\ref{eq:dm2_matt2}) agree, respectively, 
with the ones shown in eqs. (11) and (17) of \cite{Yu2016}
apart from the correction factor $c^2_{13}$ in $\alpha(\rho)$
due to non-zero $\theta_{13}$.

From eqs.~(\ref{eq:theta12_matt2}) and ~(\ref{eq:dm2_matt2}) 
we can try to estimate the expected shift 
due to the matter effect for JUNO by choosing 
some representative value of reactor neutrino energy. 
It seems reasonable to consider the neutrino energy 
corresponding to the first oscillation minimum driven by 
$\Delta m^2_{21}$ for the JUNO
baseline ($L=52.5$ km) as follows, 
\begin{eqnarray}
E^{\text{osc. min}} 
\approx 3.12
\left(\frac{\Delta m^2_{21}}{7.34\times 10^{-5}\ \text{eV}^2} \right)
\ \text{MeV}.
\end{eqnarray}

By using this energy and $\rho = 2.6$ g/cm$^3$, we expect that 
if we perform a fit to the data (which inevitably include matter effect) 
by ignoring the matter effect, 
the best fitted values of $\sin ^{2}\theta_{12} $ and $\Delta
{m}^{2}_{21}$ would be shifted as
\begin{eqnarray}
& & \left. \left( \frac{\delta( \sin^2\theta_{12})
 }{\sin^2\theta_{12}}, \frac{ \delta( \Delta m^2_{21}) }{\Delta
 m^2_{21}}
\right) \right|_{ \rm naive\ expectation}\nonumber \\[2mm]
& & 
\hspace{-1.0cm}
\simeq  
\left(
-
2\alpha (\rho) E_{\text{osc. min}} 
\, \left(\frac{\cos^2 \theta_{12}}{\cos2 \theta_{12}}\right), \ \ 
\alpha (\rho) E_{\text{osc. min}} 
\right), \nonumber \\
& & \hspace{-1.0cm}
\simeq   (-1.2,~~ 0.33)\%, 
\label{eq:shift_naive2}
\end{eqnarray}
where
\begin{eqnarray}
\frac{\delta( \sin^2\theta_{12}) }{\sin^2\theta_{12}}
&\equiv &
\frac{\sin^{2}\theta_{12}^{\text{fit}}-\sin^2\theta_{12}}{\sin^2\theta_{12}}, \\
\frac{ \delta( \Delta m^2_{12}) }{\Delta m^2_{12}} 
&\equiv &
\frac{ 
\Delta {m^2_{12}}^{\hskip -0.1cm \text{fit}} - \Delta m^2_{12}}
{\Delta m^2_{12}}.
\end{eqnarray}
In this work $\sin^{2}\theta_{12}^{\,\text{fit}}$ and 
$\Delta {m^2_{12}}^{\hskip -0.1cm \text{fit}}$ imply the mixing parameters
to be obtained by fitting the data in a $\chi^2$ analysis,
which can be performed with or without taking into matter effect.
Since one can not switch off the matter effect in a real data, we assume 
that the input data to be fitted always include the matter 
effect whereas it can be either included or neglected in the fit.
This implies that if we fit the data by using vacuum formula, 
we tend to underestimate (overestimate) the value of 
$\sin ^{2}\theta_{12} $ ($\Delta {m}^{2}_{21}$) by these percentages
using $E^{\text{osc. min}}$. 

Now let us compare this estimation with the results which can be 
obtained by fitting the data.  
As mentioned in the introduction, the naive expectation 
of the shift given in eq.~(\ref{eq:shift_naive2}) does not 
agree with the results of the full 
$\chi^2$ analysis done in \cite{Yu2016}
(see Fig. 5 of this reference)
which implies that the shift is 
\begin{eqnarray}
\left. \left( \frac{\delta( \sin^2\theta_{12})
 }{\sin^2\theta_{12}}, \frac{ \delta( \Delta m^2_{21}) }{\Delta m^2_{21}} \right) \right|_{\text{purely numerical} ~\chi^2} & & \nonumber \\[3mm]
&&  \hspace{-3.5cm} \simeq 
( -1.1,\quad  0.19)\%.  %\nonumber %
\label{eq:shift_chisq2}
\end{eqnarray}
which we confirmed also by our $\chi^2$ fit performed in a similar way as 
done in ~\cite{Abrahao:2015rba}, giving very
similar results for the fractional differences 
of ($\sin^2 \theta_{12}$, $\Delta m^2_{21}) \sim$ ( -1.2,\quad  0.20)\%.

By comparing the values given in eqs. ~(\ref{eq:shift_naive2})
and (\ref{eq:shift_chisq2}), 
we observe that while the shift for the mixing angle agrees quite well
between the one expected by naive prediction and with that 
obtained by $\chi^2$ analysis, 
we see that the shift for $\Delta {m}^{2}_{21}$ does not agree very well 
(0.33\% vs 0.19\%).
As long as $\Delta {m}^{2}_{21}$ is concerned we would need to use 
$E \sim$ 1.9 MeV to get the correct shift but this looks odd 
as this value is too close to the energy threshold for the inverse beta
decay,
and moreover, with such a energy, 
the agreement for the shift for the mixing angle gets
worse,
as mentioned in the introduction, eq. (\ref{eq:shift_naive1}).  
We have checked that taking into account higher order correction 
of matter effect would not help to make the agreement better. 

In the next subsection, we try to compute more accurately 
the shift of solar mixing parameters due to matter effect without
performing a fit to the data through a detailed numerical $\chi^2$ analysis. 

\subsection{Semi-analytic computation of the expected shift due to matter effect}
\label{subsec:computation-shift}
Based on the previous works such as \cite{Yu2016}
we can safely assume that in the presence of the (small) matter effect, 
if we try to fit the data by ignoring the matter effect, 
the quality of the fit would not be aggravated (compared to the case where 
the matter effect is correctly taken into account)
but the best fit values of the mixing parameters are simply 
shifted from the true (correct) values by some small constants
\footnote{This implies that JUNO alone can not identify 
or detect the Earth
matter effect in neutrino oscillation experimentally
unless we know {\it a priori} the true values of solar mixing parameters.}.

We try to parameterize such constants by 
dimensionless parameters $x$ and $y$ as follows, 
\begin{eqnarray}
\sin ^{2}2\theta_{12}^{\text{\,fit}} &=& \sin ^{2}2\theta_{12}
(1+x)\nonumber \\ 
\Delta m_{21}^{2\,\text{\,fit}}&=&\Delta m^2_{21}(1+y),
\label{eq:fit-shift}
\end{eqnarray}
where $x$ ($y$) is expected to take some negative (positive) 
value which implies that mixing angle (mass squared difference)
tends to be underestimated (overestimated), in agreement
with what has been obtained previously \cite{Yu2016}.
Here we consider the shift in terms of $\sin^2 2 \theta_{12}$, 
to make the computation 
simpler and then convert the shift to $\sin^2 \theta_{12}$ for
comparison.

Note that $x$ and $y$ are parameters which do not depend explicitly on
neutrino energy, so that the relations in eq.~(\ref{eq:fit-shift})
are not equivalent to that given 
in eqs.~(\ref{eq:theta12_matt3}) and (\ref{eq:dm2_matt2}). 
If the energy spectrum were monochromatic (i.e, $E=E_0$), 
we have, $x=-2\alpha(\rho) E_0$ and $y= \alpha(\rho) E_0$, so 
one naively expects that $x \approx -2y$.  For the case of 
a non-monochromatic 
spectrum, this is turned out to be not true 
due to the non-trivial energy dependence 
in the event number distribution as we will see in the analytic derivation  below. 

First let us ignore the effect of energy smearing due to 
finite energy resolution or assume that energy resolution is perfect.
This is justified as JUNO's expected energy resolution, 3\%, 
is good enough or more than necessary for solar parameter determination
so that assuming it is perfect would be reasonable for our purpose
\footnote{Let us stress that here we are interested in estimating
only the size of the shift not the associated uncertainties 
of the fitted values of the mixing parameters.}.

Then we can write down the expected event number distribution $N(E)$
at the detector as function of the neutrino energy $E$ simply 
as follows%
\begin{equation*}
N(E)=N_{0}(E)\ \tilde{P}_{ee}(E),
\end{equation*}%
where $N_{0}(E) $ is the event number distribution in the absence
of oscillation and $\tilde{P}_{ee}(E)$ is $\overline{\nu }_{e}$ 
survival probability in matter computed by eq.~(\ref{eq:p_ee-matter-approx})
with effective mixing parameters with the first order 
in matter effect given in 
eqs.~(\ref{eq:dm2_matt2}) and (\ref{eq:theta12_matt3}).

Let us expand $\tilde{P}_{ee}(E)$ considering only the first order 
in matter effect as follows,
\begin{eqnarray}
\hskip -0.5cm 
\tilde{P}_{ee}(E)  \approx P_{ee}^{\text{vac}}(E)  + 
\delta P_{ee}
\left(\frac{\delta (\sin^2 2\theta_{12})}{\sin^2 2\theta_{12}}, 
\frac{\delta (\Delta m^2_{21})}{\Delta m^2_{21}}
\right),
\end{eqnarray}
where $P_{ee}^{\text{vac}}(E)$ is the vacuum oscillation probability
which has the same form as given in eq.~(\ref{eq:p_ee-matter-approx})
but without matter effect, or explicitly, 
\begin{equation}
{P}_{ee}^{\text{vac}}(E) \ \simeq \ 
1-c^4_{13} \sin ^{2}2 \theta_{12}
\sin^{2}  \left( \frac{\Delta m^{2}_{21}L}{4E}\right) 
-  \frac{1}{2}\sin^2 2\theta_{13},
\label{eq:p_ee-vac-approx}
\end{equation}
and the second term is the correction due to matter effect, 
\begin{eqnarray}
& & \delta P_{ee}
\left(\frac{\delta (\sin^2 2\theta_{12})}{\sin^2 2\theta_{12}}, 
\frac{\delta (\Delta m^2_{21})}{\Delta m^2_{21}}
\right)
\nonumber \\
& & \hskip -1.0cm \equiv
\frac{\partial \tilde{P}_{ee}}{\partial 
(\sin^2 2\theta_{12})} \delta (\sin^2 2\theta_{12}) 
+ \frac{\partial \tilde{P}_{ee}}{\partial 
(\Delta m^2_{21})} \delta (\Delta m^2_{21}), \nonumber \\
& = & - c^4_{13}\,\sin ^{2}2\theta_{12} \sin \Delta_{21} \nonumber \\
&&\hskip -1.4cm 
\times \left[
\sin \Delta_{21}\,
\frac{\delta (\sin^2 2\theta_{12})}{\sin^2 2\theta_{12}} 
+ 2\Delta_{12}\cos \Delta_{12}\,
\frac{\delta (\Delta m^2_{21})}{\Delta m^2_{21}}
\right], 
\label{eq:function-delta-P}
\end{eqnarray}
where
$\delta (\sin^2 2\theta_{12})/ \sin^2 2\theta_{12} = -2\alpha(\rho) E$  and \\
$\delta (\Delta m^2_{21})/\Delta m^2_{21} = \alpha(\rho) E$, consistent with what is given in 
eqs.~(\ref{eq:theta12_matt3}) and (\ref{eq:dm2_matt2}).

Then the expected observed event number distribution 
in the presence of matter effects for small value of $\alpha (\rho)E$,
where $\alpha(\rho)$ is given in eq.~(\ref{eq:alpha}),
can be written as, 
\begin{align}
\hskip -0.6cm
N^{\text{obs}}(E) &\approx 
N^{\text{tot}}_0 \lambda(E) 
[{P}_{ee}^{\text{vac}}(E)+
\delta P_{ee}
(-2\alpha(\rho) E, \alpha(\rho) E)],  \nonumber \\ &
\label{nobs}
\end{align}%
where 
$N^{\text{tot}}_0$ is the expected total number of event in the absence of
oscillation, 
\begin{align}
N^{\text{tot}}_0  &= \int S(E)\cdot \sigma_{IBD}(E)dE,
\label{eq:Ntot0}
\end{align}%
and $\lambda(E)$
is the normalized event number distribution in the absence of
oscillation, defined as 
\begin{align}
\lambda (E) &\equiv S(E)\cdot \sigma _{IBD}(E)/N^{\text{tot}}_0 \, ,
\label{eq:lambda}
\end{align}%
where $S(E)$ is the reactor neutrino spectrum 
taken from \cite{Mueller:2011nm} and $\sigma _{IBD}(E)$ is 
the inverse-beta decay cross-section taken from \cite{Strumia:2003zx}. 

Now let us assume that this event number distribution can be
fitted (mimicked) by using the vacuum oscillation probability 
but with slightly shifted mixing parameters 
given in eq.~(\ref{eq:fit-shift}), i.e.
$\sin^22\theta_{12}(1+x)$ and $\Delta m^2_{21}(1+y)$,
which do not depend explicitly on neutrino energy, as if the matter effect were absent as,
\begin{eqnarray}
N^{\text{fit}}(E) \approx 
N^{\text{tot}}_0 \lambda(E) [ P_{ee}^{\text{vac}}(E)+ \delta P_{ee}(x,y)], 
\label{nfit}
\end{eqnarray}
where 
$\delta P_{ee}(x,y)$ 
has exactly the 
same functional form as given 
in eq.~(\ref{eq:function-delta-P}) but just replacing 
$\delta (\sin^2 2\theta_{12})/\sin^2 2 \theta_{12}$ and
$\delta (\Delta m^2_{21})/\Delta m^2_{21}$ by $x$ and $y$, respectively. 
Since $x$ and $y$ do not depend explicitly on neutrino energy, 
we stress that $N^{\text{\text{fit}}}(E)$ 
can not be exactly equal to $N^{\text{obs}}(E)$.  

The dimensionless shift parameters $x$ and $y$ can be obtained
by minimizing the following $\chi^{2}$ function, 
\begin{eqnarray}
\chi ^{2}(x,y) &=&
\int \left( \frac{N^{\text{obs}}(E)-N^{\text{fit}}(E)}
{\sqrt{N^{\text{obs}}(E)}}\right)
^{2}dE,  \nonumber \\
& & 
\hskip -2.6cm =
N^{\text{tot}}_0
\int 
\frac{ 
[ 
\delta P_{ee}(-2\alpha(\rho) E,\alpha(\rho) E)-
\delta P_{ee}(x,y)
]^2
}
{\tilde{P}_{ee}(E)}
\lambda(E)dE, \nonumber \\ &&
\label{chiisq}
\end{eqnarray}%
with respect to $x$ and $y$.
The sum over the discrete energy bins has been replaced 
by the integral $\int dE$ due to large statistics, as done in ~\cite{FCap2014}, 
and all the systematic parameters were ignored as they are not expected 
to be relevant as long as the computation of the shift is concerned. 

After plugging $N^{\text{obs}}(E)$ and 
$N^{\text{fit}}(E)$ from eqs.~(\ref{nobs}) and (\ref{nfit}) 
in eq.~(\ref{chiisq}), and
ignoring $\delta P_{ee}$ term in 
the denominator of the integrand of eq.~(\ref{chiisq}),
one can obtain 
\begin{eqnarray}
\hskip -0.7cm 
\chi ^{2}(x,y) 
&\propto & 
\int 
\lambda (E)\,  h(E)
\times \label{chii}  \\
&&\hskip -1.9cm 
[ (x+2\alpha(\rho) E)
\sin \Delta_{21} +2(y-\alpha(\rho) E)\Delta_{21} \cos \Delta_{21}
]^{2} \, dE \nonumber
\end{eqnarray}
where the function $h(E)$ is  given by
\begin{eqnarray}
&& \hspace{-1cm} h(E) \equiv  ( \sin^2\Delta_{21}) /  P_{ee}^{\text{vac}}(E) .
\end{eqnarray} 
The minimization condition is 
\begin{eqnarray}
\frac{\partial \chi ^{2}(x,y)}{\partial x} 
=\frac{\partial \chi ^{2}(x,y)}{\partial y}  = 0
\end{eqnarray}
which leads to coupled linear equations for the two unknowns, $x$ and $y$, 
\begin{eqnarray}
\left[
\begin{array}{cc}
a & b \\
c & d
\end{array}
\right]
\left[
\begin{array}{c}
x \\
y
\end{array}
\right]
= 
\left[
\begin{array}{c}
f \\ g
\end{array}
\right],
\label{eq:xy}
\end{eqnarray}
and $a,b,c,d,f$ and $g$ are easily derived as 
\begin{eqnarray}
a &\equiv & \int (\sin^2 \Delta_{21})\, h(E) \lambda (E)dE \approx 2.95,
\label{eq:param-a} \\
b  & \equiv & \int (\Delta_{21} \sin 2\Delta_{21} )\, h(E)
 \lambda (E)dE \, \approx 0.823,
\label{eq:param-b}  \\
c & \equiv &  b/2  \approx 0.411, \label{eq:param-c}  \\ %0.364 \\
d &\equiv & 2  \int ( \Delta_{21}^2 \cos^2\Delta_{21})
\, h(E) \lambda (E)dE \approx   1.17,
\label{eq:param-d} 
\end{eqnarray}
and
\begin{eqnarray}
\hskip -2cm f(\rho) &\equiv & - \alpha(\rho) 
\int (2 \sin^2 \Delta_{21} -\Delta_{21} \sin 2\Delta_{21}) \nonumber \\
&& \hskip 1.5cm   \times h(E)\lambda (E)EdE  \nonumber \\
& \approx &  -1.75 \times 10^{-2}  
\left( \frac{  \rho  }{2.6\  \text{g/cm}^{3}}\right) \cos^2\theta_{13} \, ,
\label{eq:param-f}
\end{eqnarray}
\begin{eqnarray}
\hskip -2cm g(\rho) &\equiv& - \alpha(\rho)
\int (\Delta_{21} \sin2\Delta_{21} -2\Delta^2_{21} \cos^2 \Delta_{21})    \nonumber \\
&& \hskip 1.5cm   \times h(E)\lambda (E)EdE  \nonumber \\
& \approx &  -3.38 \times 10^{-4}  \left( \frac{  \rho  }{2.6\  \text{g/cm}^{3}}\right) \cos^2\theta_{13} \, ,
\label{eq:param-g}
\end{eqnarray}
all in units of MeV.  The numerical values are computed using the parameters from Table~\ref{tabosc}.

Eq. \ref{eq:xy} can easily  be solved and 
the fractional shift parameters $x$ and $y$ are given as follows, 
\begin{eqnarray}
\hspace*{-1cm}
\left[
\begin{array}{c}
x \\
y
\end{array}
\right]
= 
\left[
\begin{array}{cc}
a & b \\
c & d
\end{array}
\right]^{-1}
\left[
\begin{array}{c}
f \\ g
\end{array}
\right]
% \nonumber \\
= \frac{1}{(a.d-b.c)}
\left[
\begin{array}{c}
d.f-b.g \\[1mm] a.g-c.f
\end{array}
\right].
\label{eq:lineasol}
\end{eqnarray}
Therefore 
\begin{align}
x&=\frac{\delta( \sin^22\theta_{12})
 }{\sin^2 2\theta_{12}} = \frac{d.f(\rho)-b.g(\rho)}{a.d-b.c} 
 \label{eq:th12_anal}  \\
 &   =   -6.48 \times 10^{-3}  \left( \frac{  \rho  }{2.6\  \text{g/cm}^{3}}\right) \cos^2\theta_{13}   \nonumber \\
  {\rm and} \hspace{1.25cm} & \nonumber \\[2mm]
y &=  \frac{ \delta( \Delta m^2_{21}) }{\Delta
 m^2_{21}}  =
 \frac{a.g(\rho)-c.f(\rho)}{a.d-b.c} 
  \label{eq:dmsq21_anal}\\
    &  =  ~~1.99  \times 10^{-3}  \left( \frac{  \rho  }{2.6\  \text{g/cm}^{3}}\right) \cos^2\theta_{13}.   \nonumber
\end{align}
The following points should be noted:
\begin{enumerate}
\item Since $|f| > 50\,|g|$ and a, b, c, d are all of ${\cal O}(1)$, the f terms dominate over g terms. Therefore one expects that $x \approx  -(d/c)y \sim -3 y$.
\item The size of the expected shift does not depend on 
 the total number of events or any other characteristic of the experiment apart from the baseline and the earth matter density to  the reactor(s).
\item  Apart from the evaluation of the integrals in Eqs.  (\ref{eq:param-a}) - (\ref{eq:param-g}), for  JUNO, our analysis is purely analytic. 
\item We have  also explored the sensitivity of (x,y)  to the  exact spectral shape  of the  reactor anti-neutrino  spectrum and found (x,y) to be very insensitive ($\leq 1\%$) to changes in the spectral shape caused by the different fissionable isotopes; U$^{235}$, Pu$^{239}$, Pu$^{241}$ and U$^{238}$.
\item Higher order correction of matter effects is possible though the expressions for the solutions become
more complicated. 
\end{enumerate}

Therefore, for  $\rho = 2.6$ g/cm$^3$ and  $\sin^2 \theta_{13}=0.0214$, we have  
\begin{eqnarray}
& & \left. \left( \frac{\delta( \sin^2\theta_{12})
 }{\sin^2\theta_{12}}, \frac{ \delta( \Delta m^2_{21}) }{\Delta
 m^2_{21}}
\right) \right|_{ \rm this\ work}   \nonumber \\[2mm]
& & \simeq  
( x~\frac{\cos ^{2}\theta_{12} }{\cos 2\theta_{12}},\  y), \nonumber\\[3mm] 
& &
=   (-1.1,~~0.19)\%, 
\label{eq:shift_KNP}
\end{eqnarray}
which agrees well with the shift obtained by the purely numerical $\chi ^{2}$ 
analyses done in ~\cite{Yu2016}, summarized in our eq. (\ref{eq:shift_chisq2}),
and also by  an independent numerical $\chi^2$ fit
performed in the course of this  work.

Thus, as long as the corrections due to the matter effect are small, the relationship between the true parameters and the fitted parameters are given by
\begin{align}
\sin ^{2}\theta_{12}^{\text{true}} &\approx \sin
 ^{2}{\theta^{\text{fit}}_{12}}
\nonumber  \\
& \hskip -0.7cm \times 
\left[
1- \left( \frac{d.f(\Delta \rho)-b.g(\Delta \rho)}{a.d-b.c} \right) \left( \frac{\cos ^{2}\theta_{12} }{\cos 2\theta_{12} }\right) 
 \right] 
 \label{eq:shift_theta12_ourprediction}
\end{align}
and
\begin{align}
\Delta m_{21}^{2 ~\text{true}} 
&\approx \Delta m^{2 ~\text{fit}}_{21}\left[ 1+\left( \frac{a.g(\Delta \rho)-c.f(\Delta \rho)}{a.d-b.c} \right)
  \right],
\label{eq:shift_dm2_ourprediction}
\end{align}
where $\Delta \rho \equiv ( \rho_{\text{true}}  - \rho_{\text{fit}})$, $\rho_{\text{fit}}$ is the value of the density used in the fit and $\rho_{\text{true}}$ is the actual density for the experiment/data.  Note $f(0)=g(0)=0$, so if the fit density equals the true density, the correct answer is obtained.

\begin{figure*}[h!]
\vglue -0.5cm
\begin{center}
\hspace*{1.5cm}
\includegraphics[width=0.95\textwidth]
{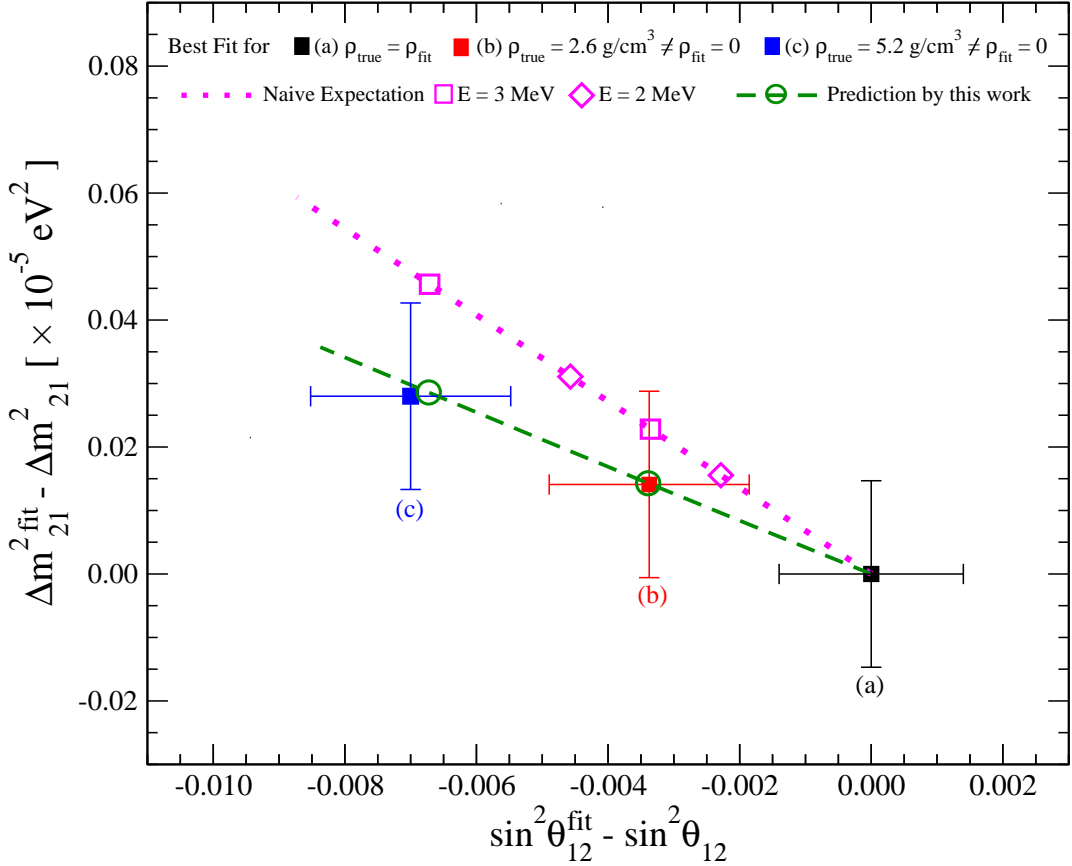}
\end{center}
\par
\vglue-0.5cm \vglue-0.5cm
\caption{
We show the best fit points 
(indicated by the filled squares with 
bars indicating 1 $\sigma$ statistical errors only)
obtained by a numerical $\chi^2$
analysis in the plane of $(\sin^2 \theta_{12}^{\text{ fit}}-\sin^2 \theta_{12})$
vs. $ (\Delta {m^{2\ {\text{fit}}}_{21}}-\Delta {m^{2}_{21}})$
for the 3 cases where 
(a) $\rho_{\text{true}} = \rho_{\text{fit}}$ (= 0 or 2.6 g/cm$^{3}$),
indicated by black color, 
(b) $\rho_{\text{true}} = 2.6$ g/cm$^{3}\ne \rho_{\text{fit}}$ = 0, 
indicated by red color
and (c) $\rho_{\text{true}} = 5.2$ g/cm$^{3}\ne \rho_{\text{fit}}$ = 0, 
indicated by blue color. 
Simultaneously in the same plot, we also show predicted shifts due to 
the matter effect obtained by naive expectation,  eq.~(\ref{eq:shift_naive2}), 
by the dashed magenta line and 
our predictions, ~eqs.~(\ref{eq:shift_theta12_ourprediction}) and 
~(\ref{eq:shift_dm2_ourprediction}), 
by the dashed green line where 
$\rho_{\text{true}}$ 
is varied continuously, from right to left, from 0 to
6.5 g/cm$^{3}$, with $\rho_\text{fit}$=0.
The two open green circles and magenta squares/diamonds correspond to the case where  
$\rho_{\text{true}}$ = 2.6 ( to the right) 
and  5.2 (to the left ) g/cm$^{3}$. For the magenta squares and diamonds, $E_\nu$=3.12 MeV and  1.9 MeV respectively, demonstrating that both shifts cannot be explained using the  same neutrino energy. The assumed true values for this plot are $\sin^2 \theta_{12}=0.304$ and $\Delta m^2_{21}=7.34 \times 10^{-5}$ eV$^2$.
}
\label{fig:allowed-regions-shift}
\end{figure*}

 For JUNO the numerical  eq.( \ref{eq:shift_KNP}) and in general the  analytic eqs. (\ref{eq:shift_theta12_ourprediction}) \& (\ref{eq:shift_dm2_ourprediction}), are  the principal results of this paper.
In order to see more clearly (visually) how the shifts predicted by 
eq.~(\ref{eq:shift_theta12_ourprediction}) and (\ref{eq:shift_dm2_ourprediction}) agree well with the results
of the numerical $\chi^2$ fit, 
we show, in Fig.~\ref{fig:allowed-regions-shift},
 the results of 
the $\chi^2$ fit, shifts obtained using these equations and the naive expectation.
In this figure we show the best fit points 
(indicated by the filled squares with 1 $\sigma$ error bars
considering only statistics for simplicity) 
obtained by $\chi^2$
analysis in the plane of $\sin^2 \theta_{12}^{\text{ fit}}
-\Delta {m^{2\ {\text{fit}}}_{21}}$
for the 3 cases where 
(a) $\rho_{\text{true}} = \rho_{\text{fit}}$, ($\Delta \rho=0$)
indicated by black color, 
(b) $\rho_{\text{true}} = 2.6$ g/cm$^{3}\ne \rho_{\text{fit}}$ = 0, ($\Delta \rho=2.6$ g/cm$^{3}$)
indicated by red color
and (c) $\rho_{\text{true}} = 5.2$ g/cm$^{3}\ne \rho_{\text{fit}}$ = 0,  ($\Delta \rho=5.2$ g/cm$^{3}$)
indicated by blue color. 
For the $\chi^2$ analysis, the probability was computed by the exact
solutions with matter effect including all order of corrections. 
%

%%%%%%%%%%%%%%%%%%%%%%%%%%%%%%%%%%%%%%%%%%%%%%%%%%%%%%%%%%%%%%
%%%%%%%%%%%%%%%%%%%%%%%%%%%%%%%%%%%%%%%%%%%%%%%%%%%%%%%%%%%%%%%%%%%%%%%%%%%%%%
%\clearpage

Simultaneously in the same plot, we also show predicted 
shifts due to the matter effect obtained by the naive formulas
in eq.~(\ref{eq:shift_naive2})
by the dashed magenta line 
and by our predictions given in 
eqs.~(\ref{eq:shift_theta12_ourprediction}) and 
~(\ref{eq:shift_dm2_ourprediction}) by
the dashed green line where 
$\rho_{\text{true}}$ 
is varied continuously, from right to left, from 0 to
6.5 g/cm$^{3}$ and 2 open green circles and magenta squares 
correspond to the case where  $\rho_{\text{true}}$ = 2.6 
g/cm$^{3}$ (right open circle/square) and 
5.2 g/cm$^{3}$ (left open circle/square) g/cm$^{3}$.
Note that some unlikely values of $\rho_{\text{true}}$ 
(significantly different from the reference value 2.6 g/cm$^{3}$)
were considered just for the sake of illustration. 

We first observe that naive predictions for the shift obtained 
by eq.~(\ref{eq:shift_naive2}) 
do not agree very well with the results of $\chi^2$ fit 
because the shift for $\Delta m^2_{21}$ for a given matter
density is always somewhat overestimated. 
On the other hand, the best fit values obtained by 
$\chi^2$ fit (indicated by the solid squares) 
agree rather well with the predictions computed 
by eq.~(\ref{eq:lineasol}) (indicated by open green circles),
especially for the reference matter density for the JUNO baseline, $\rho=2.6$ g/cm$^3$.
We believe that the small discrepancy we see for larger matter density,
namely, for the case (c) between 
the best fit obtained by the $\chi^2$ analysis (indicated by the blue
square) and the predicted one (indicated by an open green circle close
to the blue square) mainly comes from the higher order correction of the matter effect.

As was noted in ~\cite{Yu2016} and at  the beginning of Sec.~\ref{subsec:computation-shift}, for the JUNO experiment using either the vacuum or the matter oscillation probabilities gives a good fit but with slightly shifted values for the parameters  ($\sin^2 \theta_{12}$, $\Delta m^2_{21}$).  Given that this work explains the difference between these two sets for  ($\sin^2 \theta_{12}$, $\Delta m^2_{21}$) with better precision than the experimental measurements of JUNO using a nearly totally analytic method which is independent of the detailed  characteristics of the detector apart from the baseline, this work suggested an alternative analysis for JUNO.  Use the vacuum oscillation probabilities  to perform the fit and then shift these vacuum fitted values by the amounts given in this manuscript, Eq. \ref{eq:shift_theta12_ourprediction} and \ref{eq:shift_dm2_ourprediction}, evaluated at the vacuum fitted values.   These shifted vacuum fitted values for ($\sin^2 \theta_{12}$, $\Delta m^2_{21}$) should be consistent with the results using the matter oscillation probabilities and thus this provides an alternative analysis which is   a useful cross check of both analyses and if they don't agree may hint at the possibility of new physics such as non-standard neutrino interactions.

\section{Summary and Conclusions}

We have given simple, perturbative expansions for the $\sin^2 \theta$'s
and $\Delta m^2$'s  in the matter potential which can be used to
calculate the electron anti-neutrino survival probability  with
precision  more than necessary
for the JUNO experiment.  
We have shown that the maximum difference between the vacuum and matter oscillation probability occurs at the solar oscillation minimum, around 3 MeV for the JUNO experiment and has a magnitude of 3.5\% (4.0\%) including (not including) energy resolution smearing.

We have argued that the naive matter effect shift for $\sin^2 \theta_{12}$ and
$\Delta m^2_{21}$ is in conflict with what is obtained from a purely numerical $\chi^2$ analysis,  \cite{Yu2016}.
Then, we explain the apparent discrepancy between these two estimates
using a semi-analytic approaches that takes into account the shape of the reactor 
anti-neutrino spectrum and the energy dependence of the IBD cross section.  
For the detector, only the baseline and earth matter density between reactor and detector is relevant.
No other details of the experimental setup are important for the size of these shifts. 
Therefore, this paper provides independent confirmation of the size of the shifts given in \cite{Yu2016} using a totally different approach which is principally analytic and identifies the relevant and irrelevant components to these shifts, in contrast to the purely numerical $\chi^2$ analysis of \cite{Yu2016}. 

The above statements are summarized as follows:
\begin{eqnarray}
 & &  \left( \frac{\delta( \sin^2\theta_{12})
 }{\sin^2\theta_{12}}, \frac{ \delta( \Delta m^2_{21}) }{\Delta
 m^2_{21}} \right)  \nonumber \\[2mm]
& \simeq & % \hspace{+2cm}
  \left\{\begin{array}{lll}
(-1.1,~~ &0.30)\% & \text{with} ~E=3 {\rm~MeV} \\[2mm]
(-0.74, &0.21)\% & \text{with} ~E=2 {\rm~MeV}
\end{array}  \right.  \nonumber \\
& & \quad \text{using the naive expectation, eq. (\ref{eq:shift_naive1})} \nonumber  \\[3mm]
& \simeq & ( -1.1,\quad  0.19)\% \quad \text{purely numerical} ~\chi^2  \nonumber  \\
&& \quad \text{from ref.  \cite{Yu2016}, \text{given here in eq.} (\ref{eq:shift_chisq2}) } \nonumber \\[3mm]
& \simeq & ( -1.1,\quad  0.20)\% \quad \text{semi-analytic method,  } \nonumber \\
&& \quad \text{this work, eq. (\ref{eq:shift_KNP}).}  \nonumber
\end{eqnarray}
The explanation of the size of the matter effect shift for $\sin^2 \theta_{12}$ and $\Delta m^2_{21}$, using a semi-analytic method, is the main result of this paper. 

Once JUNO has real data,  it will be an interesting and useful cross check
to also perform a fit to the data by ignoring  the matter effect  in the oscillation probabilities
to obtain the solar mixing parameters, apply the shift calculated in this manuscript  
and then compare them to the ones obtained with the matter effect include the oscillation probabilities.
Assuming no new physics or analysis errors, these two measurements will be in total agreement.

\newpage

\section*{Acknowledgments}
This manuscript has been authored by Fermi Research
Alliance, LLC under Contract No. DE-AC02-07CH11359 with 
the U.S. Department of Energy, Office of Science, 
Office of High Energy Physics, and also by 
the funding and support from the European Unions Hori-
zon 2020 research and innovation programme under the
Marie \\ 
Sklodowska-Curie 
grant agreement No 690575 and No 674896.
ANK and HN thanks the NPC (Neutrino Physics Center) program of Fermilab 
for the financial support during the period when this work has started 
in 2018. HN appreciates also the hospitality of the Fermilab 
Theoretical Physics Department (TPD) where 
the final part of this work was done during 
his visit to TPD under the Summer Visitor's Program of 2019. 
HN was also supported by the Brazilian funding agency, CNPq.

\end{document}